\documentclass[showpacs,aps,floatfix,prx,reprint,superscriptaddress]{revtex4-1} 
\usepackage{xfrac}
\usepackage{amssymb}
\usepackage{braket}
\usepackage{graphicx}
\usepackage{verbatim}
\usepackage{etoolbox}
\usepackage{amsfonts} 
\usepackage{amsmath} 
\usepackage[utf8]{inputenc} 
\usepackage{mathtools}
\usepackage{soul,xcolor,colortbl}
\usepackage{setspace}
\usepackage{multirow}
\usepackage{hyperref} \hypersetup{colorlinks=true,citecolor=blue,linkcolor=blue,urlcolor=blue} 
\usepackage{bm} 
\usepackage{array}
\newcolumntype{C}[1]{>{\centering\arraybackslash}p{#1}}\usepackage{soul}
\definecolor{Gray}{gray}{0.85}

\usepackage{color} 
\usepackage{morefloats} 

\definecolor{Gray}{gray}{0.9}
\definecolor{LightCyan}{rgb}{0.88,1,1}

\def\AFLOW{{\small AFLOW}}
\def\mix{{\textrm{mix}}}
\def\bcc{{\textrm{bcc}}}
\def\hcp{{\textrm{hcp}}}
\def\ex{{\textrm{ex}}}
\def\TM{{\textrm{TM}}}
\def\Ta{{\textrm{Ta}}}
\def\Ti{{\textrm{Ti}}}
\def\Mo{{\textrm{Mo}}}

\begin{document}
\title{\Large Addressing the lattice stability puzzle in the
  computational determination of intermetallic phase diagrams}

\author{Shmuel Barzilai}
\affiliation{Department of Materials Science, NRCN, P.O.Box 9001, Beer-Sheva 84190, Israel}
\author{Cormac Toher}
\affiliation{Department of Mechanical Engineering and Materials Science, Duke University, Durham, North Carolina 27708, USA}
\author{Stefano Curtarolo}
\email[]{stefano@duke.edu}
\affiliation{Department of Mechanical Engineering and Materials Science, Duke University, Durham, North Carolina 27708, USA}
\author{Ohad Levy}
\affiliation{Department of Mechanical Engineering and Materials Science, Duke University, Durham, North Carolina 27708, USA}
\affiliation{Department of Physics, NRCN, P.O.Box 9001, Beer-Sheva 84190, Israel}

\date{\today}

\begin{abstract}
The evaluation of phase stabilities of unstable elemental phases is a long-standing problem in the computational assessment of phase diagrams. 
Here we tackle this problem by explicitly calculating phase diagrams of intermetallic systems where its effect should be most conspicuous, 
binary systems of titanium with bcc transition metals (Mo, Nb, Ta and V).  
Two types of phase diagrams are constructed: one based on the lattice
stabilities extracted from empirical data, and 
the other using the lattice stabilities computed from first principles.  
It is shown that the phase diagrams obtained using the empirical values contain clear contradictions with the experimental 
phase diagrams at the well known limits of low or high temperatures. Realistic phase diagrams, with a good agreement with 
the experimental observations, are achieved only when the computed lattice stability values are used. 
At intermediate temperatures, the computed phase diagrams resolve the controversy regarding the shape of the solvus in these systems,
predicting a complex structure with a eutectoid transition and a miscibility gap between two bcc phases.
\end{abstract}
\pacs{66.70.-f, 66.70.Df 
}

\maketitle


The systematic experimental inspection of thermodynamic properties of
alloys is usually based on phenomenological rules and metallurgical
experience. It requires considerable
efforts commonly associated with melting, casting, heat treatments,
homogenization and characterization of series of alloys, which set
practical limits on its applicability to a relatively
small part of the potential alloy space. Theoretical inspection based
on a bottom-up design strategy may enhance the experimental approach,
and ultimately replace most of it, by
providing rapid computational predictions of structural and
thermodynamic properties. The bottom-up strategy becomes viable with
the advent of powerful computational methods and
available resources. Moreover, the theoretical approach brings a
better understanding of alloy design and can reveal shortcuts to
achieve the desired alloys as opposed to trial and error
methods used in the conventional metallurgical screening.

The fundamental quantity expressing the thermodynamic stability of
alloys is the free energy. The free energy of a solid solution phase ($\phi$) in a
binary system is expressed by 
\begin{eqnarray}
\label{equation1}
  G^\phi(x_A,x_B,T)&=&x_A\cdot {^0G_A^\phi}(T) + x_B\cdot {^0G_B^\phi}(T)+ \\
  &+&{^{\mix}G^\phi} (x_A,x_B,T) + {^{\ex}G^\phi} (x_A,x_B,T), \nonumber
\end{eqnarray}
where ${^0G_A^\phi}$ and ${^0G_B^\phi}$ are the Gibbs energies of the pure elements A and B in the $\phi$ structure, 
$T$ is the absolute temperature, $^{\mix}G^\phi$  is the mixing energy of the ideal solution
\begin{equation}
{^{\mix}G^\phi}(x_A,x_B,T) = k_B T(x_A \log x_A+x_B \log x_B)  
\label{equation2}
\end{equation}
 and $^{\ex}G^\phi$ is the excess energy that represents the effects of non-ideality
\begin{equation}
{^{\ex}G^\phi}= {^{\ex}H^\phi}+ {^{\ex}G^\phi_{vib}}
\label{equation3}
\end{equation}
with 
\begin{eqnarray}
  {^{\ex}H^\phi}&=&H^\phi_{AB}-x_A\cdot{H^\phi_A}-x_B\cdot{H^\phi_B},\,\,\,\,\,{\rm and} \nonumber \\
  {^{\ex}G^\phi_{vib}}&=&{^{\ex}G^\phi_{vib,AB}}-x_A\cdot{G^\phi_{vib,A}}-x_B\cdot{G^\phi_{vib,B}}.
  \label{equation4}
\end{eqnarray}

\begin{figure*}[bht]
  \centerline{\includegraphics*[width=0.999\textwidth,clip=true]{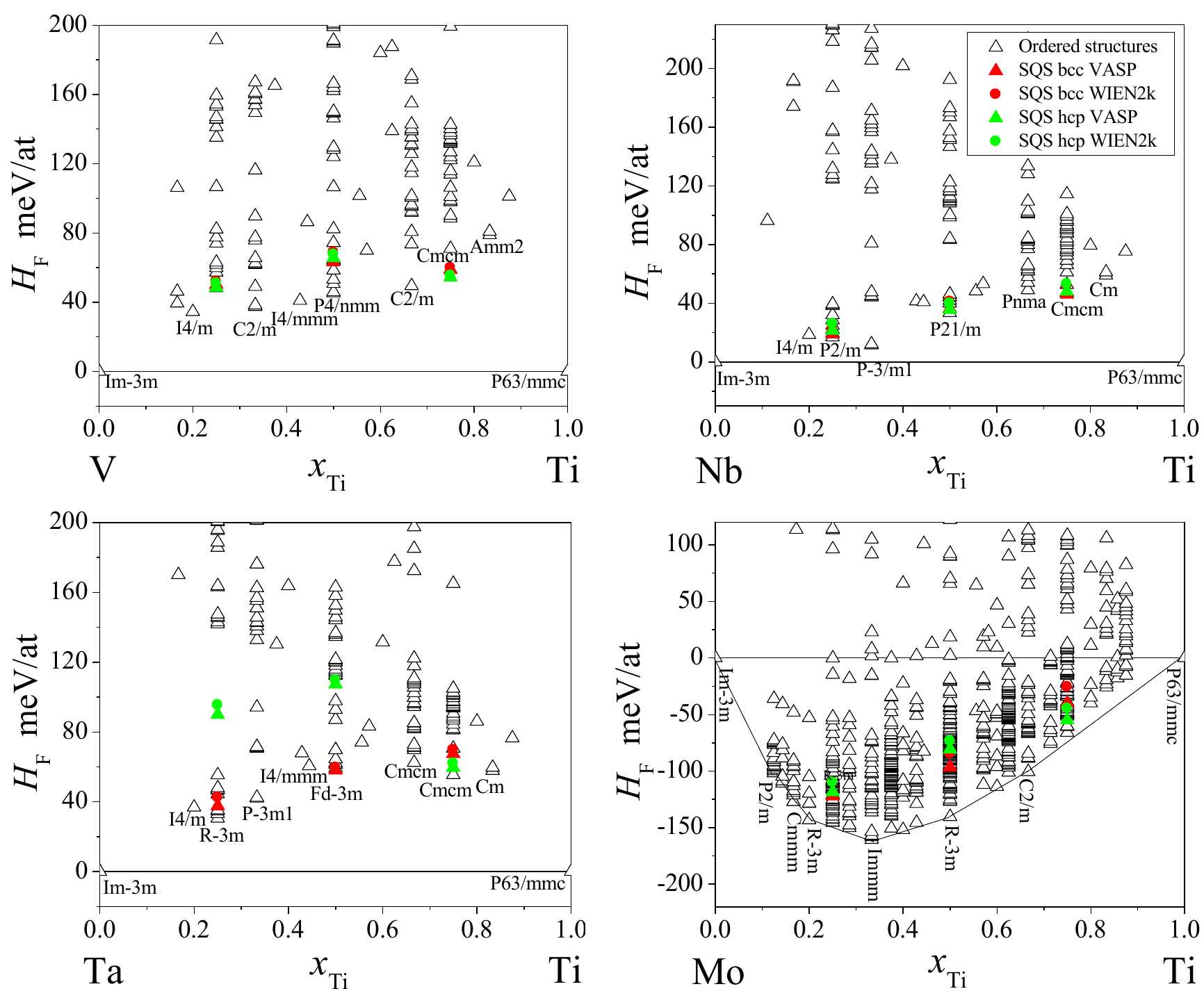}}
  \vspace{-1mm}
  \caption{\small
    Formation enthalpy of the lowest lying ordered structures and the bcc- and hcp-SQS 
    computed with respect to the ground states of the respective elements bcc-TM and hcp-Ti. }
  \label{figure1}
\end{figure*}

The phase diagram of an alloy system is determined by evaluating the Gibbs energy difference between its relevant phases, either ordered or disordered. 
The excess energy is often not available directly from simple experiments, and obtaining it requires considerable effort. 
It, therefore, usually has to be evaluated computationally.  
The Gibbs energy difference between the equilibrium and unstable phases of a single element is called {\it lattice stability}. 
This quantity is a crucial component in the construction of computational phase diagrams involving elements belonging to different lattice systems, 
e.g. an hcp-element and a bcc-element. 
In particular, it presents a conceptual problem in cases where one element is
mechanically unstable at low-temperatures in the structure of the
other, since the physical meaning of the computed
lattice stability is then unclear. 
 This problem has attracted much discussion in the literature, but has
 yet to be settled
 \cite{Grimvall_RMP_2012_lattice_instabilities,Craievich_PRL_1994_local_stability, AEKissavos-Calphad05,
   curtarolo:art15,SluiterCalphad06,Ozolins_PRL_2009_unstable_fcc_W,Usuegi_METALTRANS_2013_TiX,Palumbo_PSSB_2014_unary_phases,VanDerWalle_NCOMMS_2015_free_energy_unstable}.
 It currently presents a major impediment to the development of
 computational phase diagram databases.

Empirical estimations of lattice stabilities were developed by the
\underline{S}cientific \underline{G}roup \underline{T}hermodata
\underline{E}urope (SGTE) \cite{Dinsdale_Calphad_1991_SGTE} by direct
measurements on pure elements in their stable structures and extrapolations of activity measurements of alloys for the unstable structures.
 Alternatively, the lattice stability of the pure elements can be easily computed by {\it ab-initio} methods.  
Wang {\it et al.} \cite{curtarolo:art15} found a good agreement for the lattice stability between the computed and the empirical estimations for non-transition elements.
In these cases, only small differences up to 15 meV/atom were found. 
However, large discrepancies, most of them in the range of 80-300
meV/atom, were found for the transition metals \cite{Craievich_PRL_1994_local_stability,SluiterCalphad06}. 

Here, we consider these two approaches for assessing lattice
stabilities, one empirical and the other computational, aiming to
reveal the appropriate approach for constructing
the binary phase diagrams of mixed bcc-hcp transition metal systems. 
We chose four binary systems of the hcp-element titanium, which is
well known to be unstable in the bcc phase at low
temperatures, where a bcc solid solution is a
prominent feature at higher temperatures over the entire range of compositions. 
The empirical approach is based on the energies cited in the SGTE
database \cite{Dinsdale_Calphad_1991_SGTE}. In the computational approach, these SGTE
values are adjusted to reproduce the lattice stabilities
computed {\it ab-initio} for the two elements. 
These sets are used to construct the TM-Ti (TM=Nb, Mo, V, Ta) binary phase diagrams and thus demonstrate that the 
development of computational phase diagrams of mixed lattice
intermetallic alloy systems requires {\it ab-initio} based evaluation
of the lattice stabilities of their components.

\section{Results}

To investigate the effect of lattice stability assessment on the
predicted phase diagrams, we started by an {\it ab-initio} screening
of ordered stoichiometric structures via the high-throughput
framework \AFLOW\ \cite{aflowPAPER,curtarolo:art49}.
In addition,  the pure elements in the hcp and bcc structures and
special quasirandom structures (SQS) for the bcc \cite{Jiang_PRB_2004_SQS_bcc} and hcp
\cite{DShin-PRB06} structures were computed to estimate
the excess formation enthalpies of the corresponding solid
solutions. 

The {\it ab-initio} otal energy calculations were carried
out employing the VASP software within the AFLOW standard for
material structure calculations \cite{curtarolo:art104}. Complete information about the over 200 ordered
structures calculated per system, including initial and relaxed structures and detailed
calculation specifications, can be
obtained in the open access AFLOWLIB materials data repository
\cite{aflowlibPAPER}.
The contribution to the excess Gibbs energy at elevated
temperatures, i.e. the vibrational free energy, was
considered according to the {\small GIBBS} methodology
\cite{BlancoGIBBS2004,curtarolo:art96} in the quasiharmonic Debye model. 
The electronic and magnetic excess contributions to the Gibbs energy are
expected to be much smaller in these systems and are therefore neglected. The computed excess energies for the two phases were fitted to a sub-subregular model
\begin{eqnarray}
{^{\ex}G}=x_{\Ta}x_{\Ti}  {\Big[} {^0L_{\Ta\Ti}} + {^1L_{\Ta\Ti}} \left( x_{\Ta}-x_{\Ti}\right)+ && \nonumber \\
+ {^2L_{\Ta\Ti}} \left( x_{\Ta}-x_{\Ti}\right)^2  {\Big]} &&
\label{equation5}
\end{eqnarray}
to retrieve the Redlich-Kister coefficients that describe the excess energy over the entire composition and temperature range. 
These coefficients were used within the Thermo-Calc software
\cite{Andersson_CALPHAD_2001_THERMOCALC_DICRA} to compute two phase diagrams for each binary system, one
using the lattice stability of the pure elements taken directly from
the SGTE database \cite{Dinsdale_Calphad_1991_SGTE}, and the second using the adjusted
lattice stability values guided by the {\it ab-initio} calculations.  All
total energy calculations were carried out employing the {\small VASP}
\cite{vasp_prb1996} and the {\small WIEN2k} \cite{Schwarz_CPC_2002_WIEN2K,Cottenier_LAPW_free_book} software
packages.

The high-throughput screening of over 200 structures per system identified no stable structures in the Nb-Ti, V-Ti and Ta-Ti systems, 
in agreement with the experimental data that includes no intermetallic compounds in these systems. 
On the other hand, many ordered structures with negative formation enthalpies were found in the Mo-Ti system, of which 6 are identified as potential stable compounds.
 Figure \ref{figure1} presents these results for the various ordered structures as well as for the bcc- and hcp-SQS. 
The complete information about all the structures can be found in the
open access {\small AFLOW} materials data repository
\cite{aflowlibPAPER,curtarolo:art92,curtarolo:art104}.
 
\begin{figure*}[htb]
  \centerline{\includegraphics*[width=0.999\textwidth,clip=true]{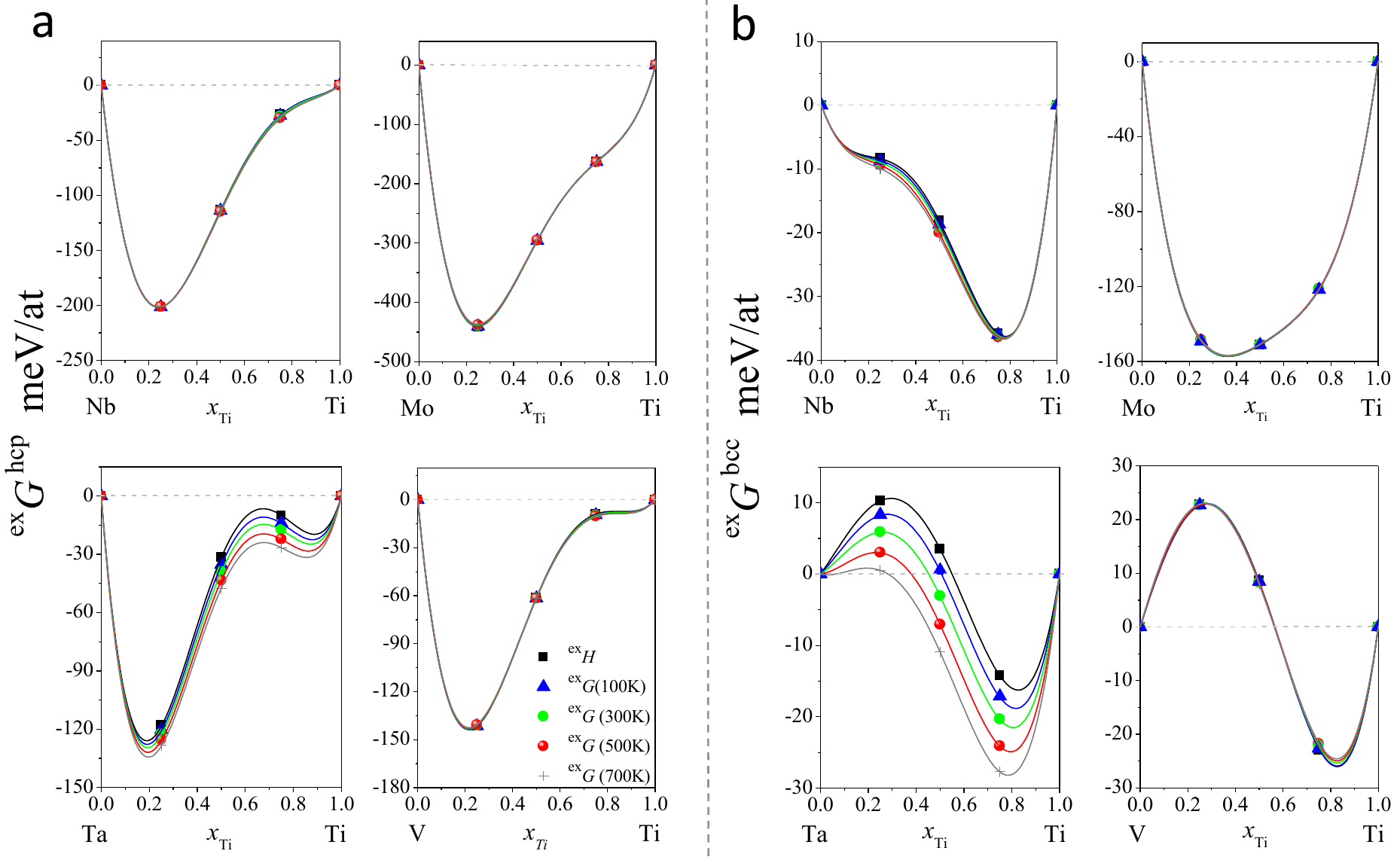}}
\vspace{-1mm}
\caption{\small
  Excess energies at 0K, ${^{\ex}H}$, and with the finite temperature contributions, ${^{\ex}G}$, computed {\it ab-initio} by the {\small GIBBS} model 
  \cite{curtarolo:art96} for the {\bf (a)} hcp structures and {\bf (b)} bcc structures. The solid lines represent the fits of the computed points to a sub-subregular model.}
\label{figure2}
\vspace{-1mm}
\end{figure*}

The Gibbs energy for a specific solid solution structure is obtained from the energies of the pure elements in the same structure, 
and the corresponding configuration and excess energies (Equation (\ref{equation1})). 
Figure \ref{figure2} shows the excess energies of the bcc and hcp structures as a function of alloy composition computed from the 
DFT total energies of the pure elements and the SQS. 
They are fitted to Redlich-Kister polynomials of the fourth degree
(Equation (\ref{equation5})) \cite{Redlich_IndsEngChem_1948} to describe the energies at all
compositions as a function of temperature. 
The DFT results obtained from the {\small VASP} and {\small WIEN2k} calculations are very close.
Both yield strong attraction for the hcp phases, but weaker attractions and in some cases even repulsive interactions for the bcc phases. 
This picture remains when the contribution of the vibrational energy is taken into account, 
the excess energies are barely affected in three of the examined systems and only a minor effect emerges in the fourth one. 
Table \ref{table1} summarizes the interaction energy coefficients between the TM and Ti atoms for both phases, as retrieved from the fitted sub-subregular model.

\begin{table}[htb]
\caption{\small
  The Redlich-Kister coefficients obtained for the sub-subregular model of the temperature dependent excess energies of the TM-Ti systems.}
\def\arraystretch{1.3}
{\small 
\begin{tabular}{|c| c |  c | c | c |}
\hline
system&phase&$^{0}L$ [meV/at]&$^{1}L$ [meV/at]&$^{2}L$ [meV/at] \\
\hline
\multirow{2}{*}{Mo-Ti}&hcp&-1185+0.009T&-1484+0.040T&-1704+0.012T \\
&bcc&-604.7+0.003T&-147.7+0.004T&-468.6+0.003T \\
\hline
\multirow{2}{*}{Ti-V}&hcp&-254.4-0.002T&705.9-0.020T&-627.6-0.002T \\
&bcc&34.2+0.003T&-242.1+0.008T&-137.6+0.034T \\
\hline
\multirow{2}{*}{Nb-Ti}&hcp&-454-0.013T&-931+0.028T&-624+0.005T \\
&bcc&-73.3-0.013T&146-0.006T&-179+0.020T \\
\hline
\multirow{2}{*}{Ta-Ti}&hcp&-133.7-0.081T&-574.4+0.042T&-865.7-0.042T \\
&bcc&9.7-0.076T&133.1+0.024T&-100.2-0.022T \\
\hline
\end{tabular}
\label{table1}
}
\end{table}

To construct a full phase diagram, it is important to define correctly the energy of the pure elements in their stable and unstable structures. 
This is especially important for solid solution systems that display strong attractions in their unstable phase. 
In our case, strong attractions were computed between the TM and Ti in
the hcp phase, thus, erroneous evaluation of the lattice stability may
prefer the unstable hcp
phase to the stable bcc phase even at the bcc rich side of the phase diagram. 
To investigate the effect of this choice we considered the two sets of ${^0}G_{\TM}$ and ${^0}G_{\Ti}$ mentioned in the introduction: 
the empirical set taken directly from the SGTE database, and the computationally guided set 
where the temperature independent
coefficients of the SGTE energies of the
unstable phases (hcp-Ta, -Mo, -Nb, -V, and bcc-Ti) were adjusted to
reproduce the {\it ab-initio} computed lattice stabilities. For Ti, the
known existence of an hcp-bcc phase transition
at 882$^\circ$C allows us to derive a temperature dependent correction.
These adjustments are summarized in Table \ref{table2}.

\begin{table}[htb]
\caption{\small
  The lattice stabilities $\Delta H_{\TM}^{\bcc-\hcp}$, of the TM elements (TM = Nb, Mo, V and Ta), 
  and $\Delta H_{\Ti}^{\hcp-\bcc}$ as computed from the SGTE values and the DFT calculations.
  The phase stability adjustments are added to the SGTE values of the ${^0}G_{\TM}$  in the hcp phase and to the ${^0}G_{\Ti}$  in the bcc phase. }
\def\arraystretch{1.5}
{\small 
\begin{tabular}{|c|c|c|c|}
\hline
elem. & \multicolumn{2}{c|}{$\Delta H_{\TM}^{\bcc-\hcp}$ or $\Delta H_{\Ti}^{\hcp-\bcc}$} & Adjustment \\
         & \multicolumn{2}{c|}{eV/at (kJ/mol)} &  eV/at (kJ/mol) \\
\hline
~~ & SGTE & DFT &~~ \\ 
\hline
Ta & -0.124~(-12)~&~-0.278~(-27)~&~0.154~(15) \\
\hline
Mo &  -0.118~(-11.5)&-0.431~(-41.8)&0.312~(30.3) \\
\hline
Nb& -0.103~(-10)~&~-0.297~(-28.8)& 0.194~(18.9) \\
\hline
V & -0.041~(-4)~&-0.253~(-24.6)& 0.212~(20.6)  \\
\hline
Ti& -0.071~(-6.8)~&-0.11~(-10.6)& 0.039-3.377·10$^{-5}$T \\
~~&~~~~~~~~~~~~~~~&~~~~~~~~~~~~&~~~~(3.8-0.00329T) \\
\hline
\end{tabular}
\label{table2}
}
\end{table}

\begin{figure*}[htb]
  \centerline{\includegraphics*[width=0.999\textwidth,clip=true]{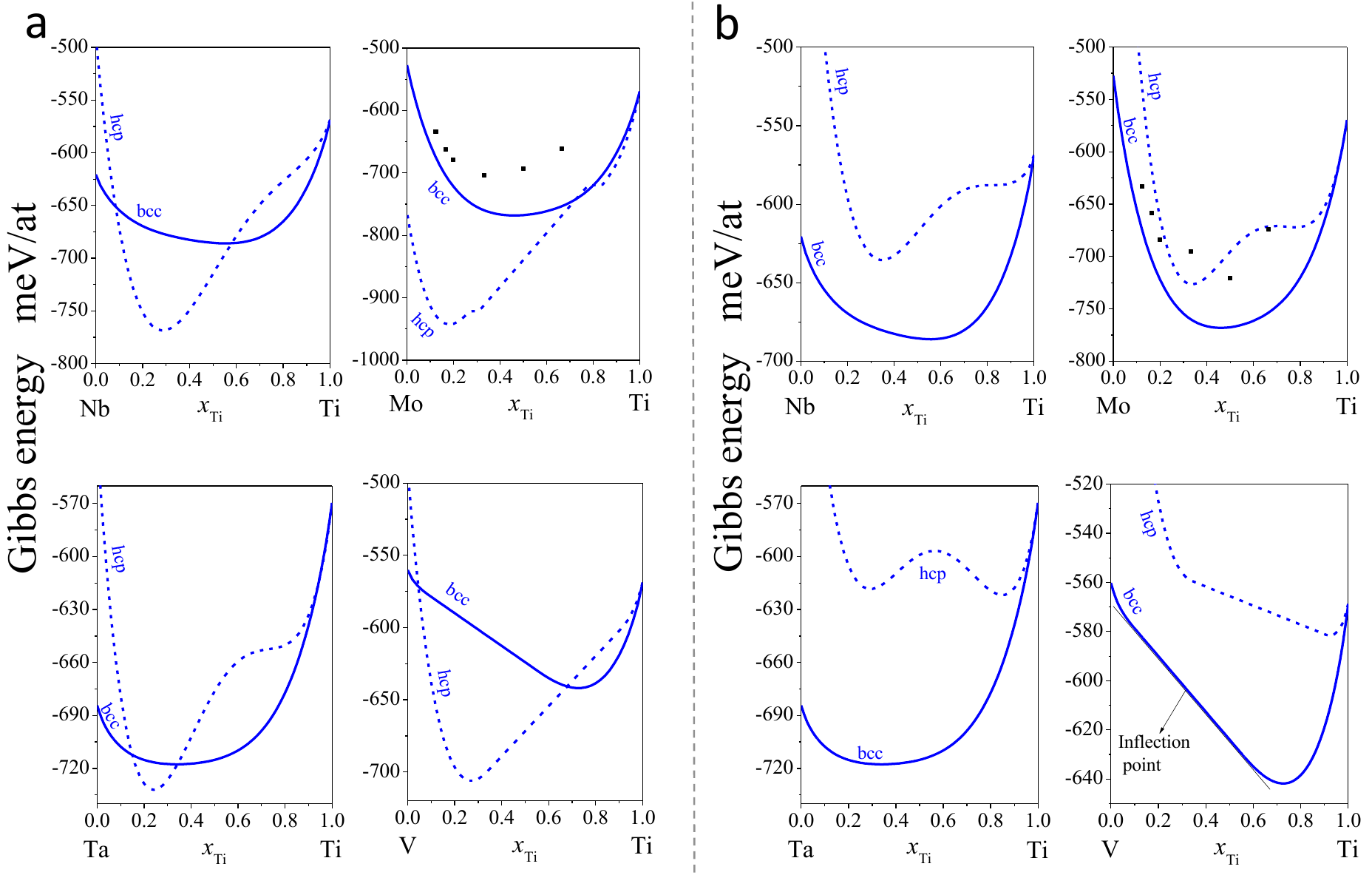}}
  \vspace{-1mm}
  \caption{\small
    The computed solid solution Gibbs energies for the hcp (dash lines) and bcc (solid lines) phases in the 
    TM-Ti systems at 882$^\circ$C, the hcp-bcc transition temperature of titanium; {\bf (a)} using
    the empirical set of ${{^0}G_{\TM}}$ and ${{^0}G_{\Ti}}$ from the SGTE database, and {\bf (b)} using the set of DFT adjusted lattice stabilities.}
  \label{figure3}
\end{figure*}

\begin{figure*}[htb]
  \centerline{\includegraphics*[height=1.25\textwidth,clip=true]{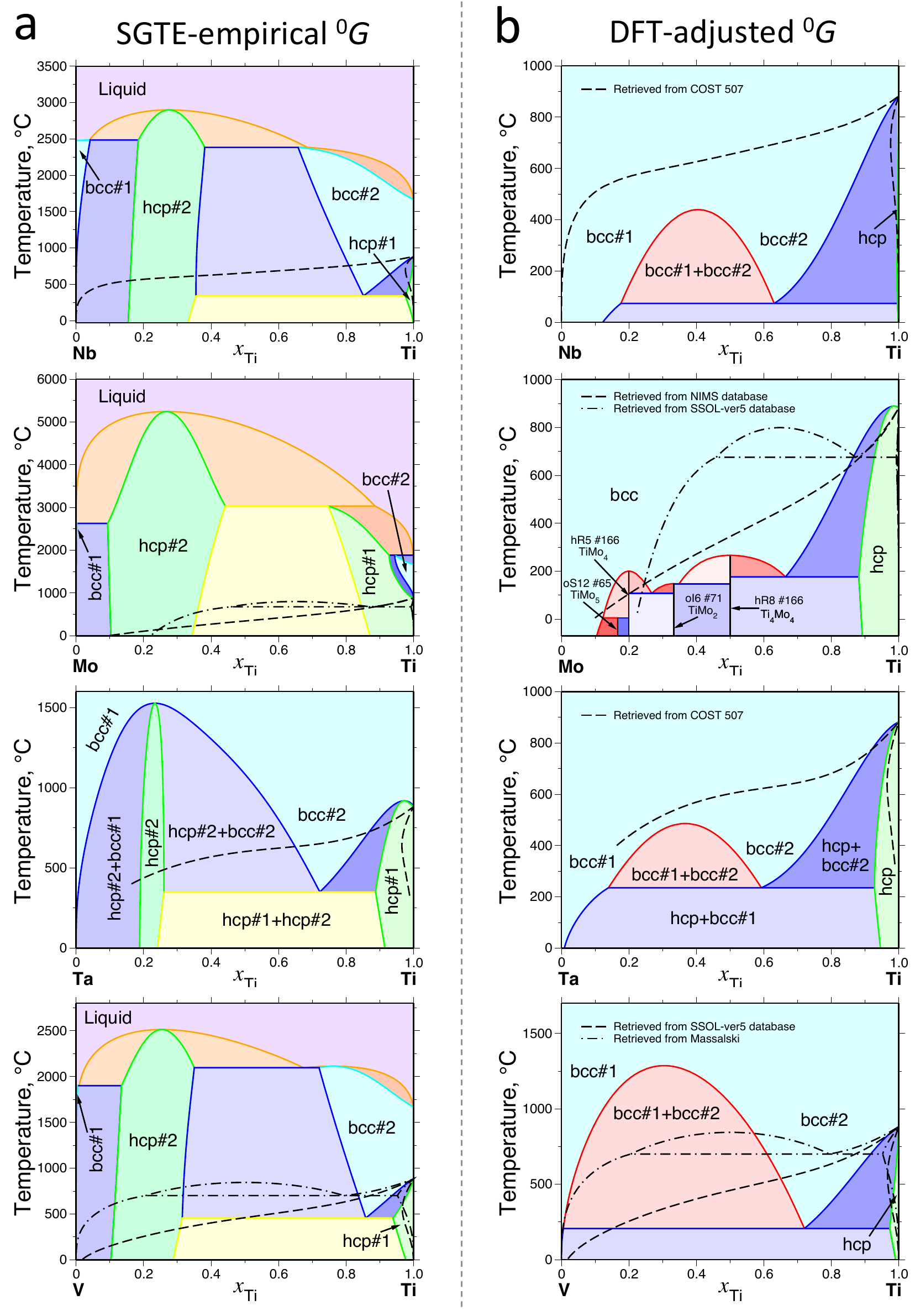}}
  \vspace{-2mm}
  \caption{\small
    The computed TM-Ti phase diagrams based on {\bf (a)} the empirical set of ${{^0}G_{\TM}}$ and ${{^0}G_{\Ti}}$ 
    from SGTE database, and {\bf (b)} DFT adjusted lattice stabilities.
    The dashed and dotted lines represent experimentally assessed diagrams \cite{ASMAlloyInternational,Massalski,Thermocalc_software,PST_Group_NIMS}. 
    DFT adjustment greatly improve predictions with respect to the experimental results.}
  \label{figure4}
\end{figure*}

As can be seen from Table \ref{table2}, both approaches show that the
bcc structure is more stable for the TM elements and the hcp structure is
more stable for Ti, in agreement with the experimental and
computational data \cite{curtarolo:calphad_2005_monster,ASM_MoTi_TaTi_VTi_NbTi,curtarolo:art87}.
However, in all of these cases the stability level computed by the DFT is higher. 
The difference in lattice stability is quite small for Ti ($\sim$0.04 eV/at), but significant for the bcc-elements.
These differences would be of little effect in ideal systems, where the attractions between the elements are relatively small. 
But, as shown in Figure \ref{figure2}, in the cases examined here, the interaction between the TM and Ti is strong, especially in the hcp phase. 
Figures \ref{figure3} and \ref{figure4} show the thermodynamic
properties of the TM-Ti systems computed with the empirical set of
${^0}G_{\TM}$ and ${^0}G_{\Ti}$, taken directly from
the SGTE (\ref{figure3}a and \ref{figure4}a) and with the
computationally guided set (\ref{figure3}b and \ref{figure4}b). Figure
\ref{figure3}  presents
the Gibbs energies of the TM-Ti alloys in the hcp and the bcc structures at 882$^\circ$C. 
It is well established that above this transition temperature of Ti from hcp to bcc, bcc is the stable phase for the entire concentration range of these alloy systems
\cite{ASM_MoTi_TaTi_VTi_NbTi,
  Fuming_MST_1989_Ti_50V,Murray_BULL_ALLOYS_1981_MoTi,ASMAlloyInternational,Massalski,Summers-Smith_JIM_1952_TaTi,
  Maykuth_TiW_TiTa_1953,Budberg_NAUK_1967_TiTa,Budberg_NAUK_1967_TiTa_translation,COST_alloy_database,Chernov_PIC3rd_1982_Ti_Alloys,Thermocalc_software,PST_Group_NIMS}. 
However, using the SGTE values for ${^0}G_{\TM}$ and ${^0}G_{\Ti}$ we find (Figure \ref{figure3}a) that the hcp solid 
solutions are more stable than the bcc solutions near 20 at \% of Ti.
This contradiction disappears using the computationally adjusted ${^0}G_{\TM}$ and ${^0}G_{\Ti}$ (Figure \ref{figure3}b). 

The binary phase diagrams for these TM-Ti systems were computed using
the Thermo-Calc software, with the excess energies from Table
\ref{table1} and the two sets of ${^0}G_{\TM}$ and ${^0}G_{\Ti}$
(Figure \ref{figure4}). Here again, unreasonable results are obtained
with the empirical set. As can be seen in Figure \ref{figure4}a, a
stable hcp compound emerges around 20 at\% of Ti. For the Mo-Ti
system, which requires the largest DFT adjustment (addition of 0.312
eV/at to ${^0}G_{\Mo}(\hcp)$  of the SGTE) the hcp compound seems to
be stable up to 5000$^\circ$C.  Obviously, this result is wrong.

The computational set of ${^0}G_{\TM}$ and ${^0}G_{\Ti}$ leads to
realistic results, as shown in Figure \ref{figure4}b. For the Mo-Ti,
Ta-Ti and Nb-Ti systems, the computed phase diagrams reproduce the known
structural features of the experimental phase diagrams at temperatures
above 500$^\circ$C
\cite{Murray_BULL_ALLOYS_1981_MoTi,ASMAlloyInternational}, i.e. a stable bcc solid solution over most
compositions and a narrow bcc-hcp phase separation at the Ti-rich
side. At low temperatures, the wide bcc-hcp phase separation is also
reproduced in the Ta-Ti, Nb-Ti and V-Ti systems, and unobserved new
compounds are predicted in the Mo-Ti system below 200$^\circ$C. These
predicted structures likely escaped detection until now due to the
slow diffusion toward equilibrium at those low temperatures. At
intermediate temperatures, the characteristics of the solvus between
the phase separation regime and the solid solution region support
those experimental reports \cite{Murray_BULL_ALLOYS_1981_MoTi,ASMAlloyInternational} that find a phase
separation hump between two stable bcc phases defined by monotectoid
points. The monotonic solvus reported by other studies
\cite{ASMAlloyInternational,Massalski} should therefore be replaced by this more complex
structure.

In the Ti-V system, the predicted bcc phase separation hump has a
critical temperature of $\sim$1200$^\circ$C and a monotectoid
temperature at $\sim$200$^\circ$C. One version
of the experimental phase diagram shows the same features but with a
critical temperature of $\sim$850$^\circ$C and a monotectoid
temperature at $\sim$675$^\circ$C, while another
version reports a monotonically decreasing solvus temperature with
increasing vanadium concentration \cite{ASMAlloyInternational}. The calculated phase
diagram clearly favors the complex solvus
structure. Its features are, however, quite sensitive to the
interaction between the pure elements. A reduction of merely 5 meV/at
of the attractive and repulsive interactions at
700K would eliminate the inflection point
shown in Figure \ref{figure3}b and reduce the critical temperature
from 1200$^\circ$C to the experimental value
of 850$^\circ$C. Such a reduction may be achieved by a more accurate modeling 
of the vibrational energy, rather than the quasiharmonic Debye model used here.

\section{Conclusions}\label{conclusions}

The determination of phase diagrams requires the evaluation of the free energy differences between multiple phases of elements, 
compounds and solid solutions based on a few lattices. This requirement leads to the problem of assigning definite values of
energy to unstable phases of the constituent elements, which is considered to preclude the direct use of lattice stabilities derived for these
structures by {\it ab-initio} electronic structure calculations. 
Attempts to resolve this problem, such as the current {\small CALPHAD} practice, are usually based on extrapolations that aim to give a 
smooth continuation of the Gibbs energy through the unstable range of alloy compositions of the investigated system. 
The assignment of the extrapolation is highly non-unique and is usually regarded as devoid of physical content but merely provides 
a convenient description of multi-component alloys. Moreover, its
consistency with established experimental data cannot be always
guaranteed and may lead to errors in the assessment of the phase diagram.

Here, we examine four binary systems of transition metal alloys, TM-Ti systems with TM= Mo, Nb, Ta and V, in which the problem of unstable phases of the elements arises. 
We assess the phase diagrams of these systems using two approaches for
the lattice stabilities of their respective constituents, one in which
they are derived from
DFT calculations and 
another where the empirical SGTE values are used without further adjustments. 
The resulting phase diagrams clearly demonstrate that attempts to mix
computational results for the unknown properties of the solid
solutions with an
empirical assessment of the lattice stabilities of the elements lead to phase diagrams that badly contradict established experimental data. 
On the other hand, the use of the {\it ab-initio} lattice stability values produces phase diagrams that agree well with the available observations and extends them to 
temperature regimes where experimental data is lacking. In the four cases we examined, it is predicted that the transition from high-temperature solid solutions to
 low-temperature phase separation is characterized by a complex solvus with a miscibility gap between two bcc phases.

{\small
\section{Acknowledgements}

C.T and S.C acknowledge partial support by DOD-ONR (N00014-13-1-0635, N00014-11-1-0136, N00014-09-1-0921). 
The {\small AFLOW} consortium would like to acknowledge the Duke University - Center for Materials Genomics and the CRAY corporation for computational support.
}

\newcommand{\Ozolins}{Ozoli\c{n}\v{s}}

\end{document}